\documentstyle{article}
\title {\vspace*{-2.1cm}\hspace*{4.4cm} {\large gr-qc/9405027,
IMAFF-RC-04-94, CGPG-94/5-2}\\ \bf \vspace*{1.2cm} Wormholes as Basis
for the Hilbert Space in Lorentzian Gravity}
\author{\vspace*{.2cm} \\Guillermo A. Mena Marug\'an \vspace*{.7cm} \\
Instituto de Matem\'aticas y F\'{\i}sica Fundamental,\\C.S.I.C., Serrano 121,
28006 Madrid, Spain,\vspace*{.3cm}\\ and \vspace*{.3cm}\\
Center for Gravitational Physics and Geometry, Dpt. of Physics,\\
Pennsylvania State University, University Park, PA 16802, USA.\vspace*{.5cm}
\\}
\date{\empty}

\begin{document}
\renewcommand{\thesection}{\Roman{section}}
\renewcommand{\theequation}{\arabic{section}.\arabic{equation}}
\maketitle
\large
\setlength{\baselineskip}{.825cm}

\begin{center}
{\Large {\bf Abstract}}
\end{center}

We carry out to completion the quantization of a Friedmann-Robertson-Walker
model provided with a conformal scalar field, and of a Kantowski-Sachs
spacetime minimally coupled to a massless scalar field. We prove that the
Hilbert space determined by the reality conditions that correspond to
Lorentzian gravity admits a basis of wormhole wave functions. This result
implies that the vector space spanned by the quantum wormholes can be
equipped with an unique inner product by demanding an adequate set of
Lorentzian reality conditions, and that the Hilbert space of wormholes
obtained in this way can be identified with the whole Hilbert space of
physical states for Lorentzian gravity. In particular, all the normalizable
quantum states can then be interpreted as superpositions of wormholes.
For each of the models considered here, we finally show that the physical
Hilbert space is separable by constructing a discrete orthonormal basis of
wormhole solutions.

\vspace*{.8cm}
\newpage

\section {Introduction}

Classical wormholes are solutions to the Euclidean equations of General
Relativity that connect two asymptotic regions of large three-volume by a
throat [1-4]. This type of instanton may exist only for matter fields
that allow the Ricci tensor to have negative eigenvalues [5,6]. Therefore,
classical wormholes cannot play a fundamental role in physics, because they
are ruled out for generic kinds of matter content. However, from the
quantum mechanical point of view, wormholes can be realized as physical
states with an appropriate asymptotic behavior for large three-geometries,
such that they can be interpreted as tubes connected to a region of spacetime
with large three-volume [7]. As opposed to the situation found in classical
gravity, quantum wormholes may exist regardless of the properties of the
matter fields.

In 1990, Hawking and Page proposed to characterize the quantum wormholes
as the wave functions that decrease exponentially in the limit of large
three-geometries and present no singularities for finite matter fields
when the three-geometry is regular or collapses to zero because of an
ill-defined slicing of the spacetime [7]. These conditions on the quantum
wormhole states were reformulated by Garay in an adequate language for the
complex path integral approach to General Relativity. Garay proved that
(at least in some minisuperspace models) the wormhole wave functions can be
defined as path integrals over Euclidean manifolds that match asymptotic large
three-geometries with fixed gravitational momenta and no gravitational
excitations [8].

Quantum wormholes can significantly affect the low energy physics in our
Universe, because they allow field interactions between distant points of the
spacetime [1]. Assuming that the space of wormholes admits a Hilbert
structure, it has been shown in the literature that the existence of quantum
wormholes may alter the values of the effective physical constants of nature
and, in particular, explain why we observe a vanishing cosmological constant
[9-10]. The conjecture that wormholes form a Hilbert space has, nonetheless,
never been proved in full gravity, although it has been demonstrated in a
number of minisuperspace models for which one can explicitly find a complete
set of wormhole solutions [11]. In these minisuperspaces, the Hilbert
structure of the wormhole wave functions was obtained by restricting to the
space of physical states an inner product that can be introduced in a natural
way in the corresponding unconstrained models [11]. In more general
gravitational systems, however, one should not expect the physical states to
be normalizable with respect to any inner product defined before imposing the
first-class constraints of the model. A more rigorous prescription is
therefore needed in order to fix the Hilbert structure in the space spanned
by the wormhole solutions. In a recent paper [12], we argued that such a
structure can in fact be uniquely determined by imposing a set of Lorentzian
reality conditions in the quantum theory, i. e., adopting Ashtekar's
proposal for the non-perturbative quantization of systems with constraints
[13,14].

We recall that reality conditions are adjointness relations between
observables\footnote{We understand observable to refer to operators that
commute with all the first-class constraints.} of the quantum theory that
capture the complex conjugation relations between the classical variables
of the system [13,14]. According to Ashtekar, these complex conjugation
relations must be promoted under quantization into an abstract
$\star$-involution of operators [13,15]. Once one has chosen a
representation for the quantum theory, what includes the selection of a
(possibly over-)complete set of elementary operators that is closed under
commutation relations and under the introduced $\star$-involution [13,14],
the physical inner product in the space of solutions to all quantum
constraints (the physical states) is fixed by imposing the
$\star$-relations between a sufficiently large number of observables as
adjointness requirements (reality conditions) [16]. In other words, given
a representation for the quantum theory, the Hilbert space of physical states
is entirely determined by the set of reality conditions imposed on the
system.

As a consequence, if the space of wormhole solutions admits in a certain
representation a Hilbert structure that can be found by demanding
Lorentzian reality conditions, as we propose [12], such Hilbert space of
wormholes has then to be a subspace of the Hilbert space of the Lorentzian
gravitational model in the adopted representation.
Moreover, if we assume that the vector space spanned by the wormhole wave
functions is stable under the action of the observables and
restrict our attention to irreducible representations, we conclude
that the Hilbert space of wormholes can be identified with that of the
considered Lorentzian model [12]. In particular, this implies that the
quantum wormhole solutions provide a basis for the Hilbert space of the
Lorentzian theory.

The above statements were rigorously proved in Ref. [12] for the case of a
Friedmann-Robertson-Walker (FRW) spacetime in the presence of a minimal
massless scalar field. Our aim in this work is to present a similar proof
for the other two minisuperspace models for which we know a complete set of
wormhole wave functions, namely, a FRW model provided with a conformal scalar
field [8,11] and a Kantowski-Sachs (KS) spacetime minimally coupled to a
massless scalar field [8,17].

The outline of the rest of this paper is as follows. In Sec. II we quantize to
completion the conformally coupled FRW spacetime. We find the Hilbert space
of physical states determined by the Lorentzian reality conditions and show
that it is spanned by a set of wormhole wave functions. The quantization of
the KS model is carried out in Sec. III. For this minisuperspace, we prove in
Sec. IV that the physical Hilbert space admits a basis of
generalized wormhole solutions, which are asymptotically damped for large
three-geometries but fail to satisfy the requirement of regularity when the
three-geometry degenerates. In Sec. V we consider other choices of basis
of wormholes for the Hilbert space of the KS model. In particular, we show
that there exists a discrete basis formed by normalizable states. Finally, we
summarize the results in Sec. VI.

\section {FRW Model}

Let us study first the case of a homogeneous scalar field conformally
coupled to a FRW spacetime. The FRW metric can be written in the form [8]
\begin{equation}ds^2=\frac{2G}{3\pi}q^2(t)
\left(-N^2(t)dt^2+d\Omega_3^2\right),\end{equation}
where $N(t)$ and $q(t)$ are, respectively, the rescaled lapse function and
the scale factor of the model, $G$ is the gravitational constant, and
$d\Omega_3^2$
is the metric of the unit three-sphere. Since the line element in (2.1)
depends only on the square of $q$, we will restrict the scale factor to be
positive, $q>0$, so that each Lorentzian
metric be considered once.
Redefining then the scalar field $\Phi(t)$ as [8]
\begin{equation} \Phi(t)=\sqrt{\frac{3}{4\pi G}}\,\frac{\chi(t)}{q(t)},
\end{equation}
we arrive at the following expression for the Hamiltonian constraint
of the minisuperspace model:
\begin{equation} H\equiv \frac{1}{2}\left(q^2+\Pi_q^2-\chi^2-\Pi_{\chi}^2
\right)=0\;\;\;\;\;(q>0),\end{equation}
with $\Pi_q$ and $\Pi_{\chi}$ the momenta canonically conjugate to $q$ and
$\chi$.

Apart from the restriction $q>0$, the above Hamiltonian is
exactly the difference of the Hamiltonians of two harmonic oscillators.
It therefore seems natural to adopt as representation space for the quantum
theory the vector space of complex
functions $\Psi$ on $I\!\!\!\,R^+\times I\!\!\!\,R$ spanned by the basis
\begin{equation} \tilde{\Psi}_{nm}(q,\chi)=\varphi_n(q)\varphi_m(\chi)\;
\;\;\;(q\in I\!\!\!\,R^+,\;\chi\in I\!\!\!\,R),\end{equation}
with $\varphi_n(x)$ the harmonic-oscillator wave functions,
\begin{equation} \varphi_n(x)=\frac{(-1)^n}{\sqrt{2^nn!\sqrt{\pi}}}\,
e^{\frac{x^2}{2\;}}\frac{d^n\;}{dx^n}(e^{-x^2}),\;\;\;\;n=0,1,...
\end{equation}
As our complete set of elementary operators we will select
the analogue of the annihilation and creation operators of the two harmonic
oscillators:
\begin{equation}
\hat{a}_{\alpha}\Psi=\frac{1}{\sqrt{2}}(\alpha+\partial_{\alpha})\Psi,\;\;\;
\hat{a}_{\alpha}^{\dagger}\Psi=\frac{1}{\sqrt{2}}(\alpha-\partial_{\alpha})
\Psi,\end{equation}
where the variable $\alpha$ denotes both $q$ and $\chi$, and we have let
$\hbar=1$. These operators represent the following complex
functions on phase space:
\begin{equation}a_{\alpha}=\frac{1}{\sqrt{2}}(\alpha+i\Pi_{\alpha}),\;\;\;
a_{\alpha}^{\dagger}=\frac{1}{\sqrt{2}}(\alpha-i\Pi_{\alpha}),\end{equation}
They form [18] a closed algebra under commutation relations, the
only non-vanishing commutators being $[\hat{a}_{\alpha},\hat{a}_{\alpha}^
{\dagger}]=1$ ($\alpha=q$ or $\chi$). On the other hand, recalling
that $q$, $\chi$, $\Pi_q$, and $\Pi_{\chi}$ are real in Lorentzian
gravity, it is easy to check that the algebra of basic operators
given by (2.6) is also closed under the $\star$-involution that is
induced by the complex conjugation relations between classical variables,
namely, $\hat{a}_{\alpha}^{\star}=\hat{a}_{\alpha}^{\dagger}$.

In the constructed representation, the Hamiltonian constraint (2.3)
can be rewritten as the Wheeler-DeWitt equation:
\begin{equation} \hat{H}\Psi(q,\chi)\equiv\left(\hat{a}_q^{\dagger}
\hat{a}_q-\hat{a}_{\chi}^{\dagger}\hat{a}_{\chi}\right)\Psi(q,\chi)=
\frac{1}{2}\left(q^2-\partial_q^2-\chi^2+\partial_{\chi}^2\right)\Psi(q,\chi)
=0.\end{equation}
Hence, the physical states of the system turn out to be linear combinations of
the wave functions
\begin{equation}\Psi_n(q,\chi)=\tilde{\Psi}_{nn}(q,\chi)=\varphi_n(q)
\varphi_n(\chi),\;\;\;\; n=0,1,...\end{equation}

{}From the expression of the Hamiltonian $\hat{H}$ in eq. (2.8) one
can prove that the operators
\begin{equation} \hat{N}_q=\hat{a}^{\dagger}_q\hat{a}_q,\;\;\;\hat{N}_{\chi}=
\hat{a}^{\dagger}_{\chi}\hat{a}_{\chi},\;\;\;\hat{J}_+=\hat{a}^{\dagger}_q
\hat{a}^{\dagger}_{\chi},\;\;\;\hat{J}_-=\hat{a}_q\hat{a}_{\chi},
\end{equation}
provide an over-complete set of observables for the model [13].
These observables [18] form a closed algebra under commutation relations, with
non-vanishing commutators given by
\begin{equation} [\hat{N}_{\alpha},\hat{J}_+]=\hat{J}_+,\;\;\;
 [\hat{N}_{\alpha},\hat{J}_-]=-\hat{J}_-\;\;\;\; (\alpha=q\;\,{\rm or}\;\,
\chi),
\end{equation}
\begin{equation}
[\hat{J}_+,\hat{J}_-]=-\hat{N}_q-\hat{N}_{\chi}-1.\end{equation}
The relations $\hat{a}_{\alpha}^{\star}=\hat{a}_{\alpha}^{\dagger}$ imply
on the other hand that
\begin{equation}\hat{N}_q^{\star}=\hat{N}_q,\;\;\;\;\hat{N}_{\chi}^{\star}=
\hat{N}_{\chi},\;\;\;\;\hat{J}_+^{\star}=\hat{J}_-,\end{equation}
while the Hamiltonian constraint (2.8) demands that the operators
$\hat{N}_q$ and $\hat{N}_{\chi}$ coincide on the physical states.
Therefore, the Lorentzian
reality conditions amount to require that, on the space of physical
states, $\hat{N}_{\chi}$ (or $\hat{N}_q$)  be self-adjoint and $\hat{J}_-$
be the adjoint of $\hat{J}_+$. We expect the inner product in such a space
to adopt the generic expression
\begin{equation} <\Upsilon,\Psi>=\int_{I\!\!\!\,R^+}dq\int_{I\!\!\!\,R}d\chi\;
g(q,\chi)\Upsilon^{\ast}(q,\chi)\Psi(q,\chi),\end{equation}
where the symbol $\ast$ denotes complex conjugation,
$g(q,\chi)$ is a measure to be determined by imposing the reality
conditions stated above, and
\begin{equation}
\Psi(q,\chi)=\sum_{n=0}^{\infty}f_n\varphi_n(q)\varphi_n(\chi),\;\;\,
\Upsilon(q,\chi)=\sum_{n=0}^{\infty}h_n\varphi_n(q)\varphi_n(\chi),\;\;\,
f_n,h_n\in /\!\!\!\!\,C.\end{equation}

Taking into account the asymptotic behavior of
the harmonic-oscillator wave functions $\varphi_n$ at infinity and that
$\varphi_{2p+1}(q=0)=0$ for all non-negative integers $p$, a straightforward
calculation shows that the Lorentzian reality conditions are satisfied if and
only if $g(q,\chi)$ is constant. Letting thus $g(q,\chi)=1$, we arrive at the
inner product
\begin{equation} <\Upsilon,\Psi>=\int_{I\!\!\!\,R^+}dq\int_{I\!\!\!\,R}d\chi
\;\Upsilon^{\ast}(q,\chi)\Psi(q,\chi)=\frac{1}{2}\sum_{n=0}^{\infty}
h_n^{\ast}f_n.\end{equation}
On the right hand side of this equation we have substituted eq. (2.15)
and used that
\begin{equation} <\Psi_n,\Psi_m>=\frac{1}{2}\delta_{nm},\end{equation}
with $\delta_{nm}$ the Kronecker delta (see eqs. (2.5) and (2.9)).
The physical Hilbert space of the model is then isomorphic to
$l^2$, the space of square summable sequences.

Notice that, in the representation that we have chosen, the Hilbert space of
the Lorentzian theory is spanned by the set of wave functions (2.9). These
functions correspond in fact to quantum wormhole states, because they are
exponentially damped for large three-geometries ($q\gg 1$), present no
singularities in the region $\{ q>0,\phi\in I\!\!\!\,R\}$, and have a
well-defined limit when $q$ tends to zero, i.e., when the three-geometry
degenerates. Hence, in agreement with our discussion in the Introduction, the
space spanned by the wormhole wave functions (2.9) does not only admit a
Hilbert structure, which can be selected by imposing an adequate set of
Lorentzian reality conditions, but, in addition, the Hilbert space of
wormholes determined in this way turns out to coincide with that of the
Lorentzian model under study.

The different elements of the basis (2.9) are described by the wormhole
parameter $n$, which is the eigenvalue reached by the observable
$\hat{N}_{\chi}$ (or $\hat{N}_q$) in these wormhole solutions. Let us
also note that, given eq. (2.17), the wormhole wave functions
$\sqrt{2}\Psi_n(q,\chi)$ ($n=0,1,...$) provide an orthonormal basis
for the physical Hilbert space, which results then in being
separable.

The wormhole states (2.9) were first considered by Garay in Refs. [8,11].
Following different arguments, he also found the inner product (2.16)
in the space of quantum wormholes, except for that he allowed the scale factor
$q$ to run over the whole real axis [11]. Since extending the
domain of definition of q in (2.16) to $I\!\!\!\,R$ only changes the right
hand side of that equation by a factor of 2, the Hilbert space of wormholes
determined in Ref. [11] is nonetheless isomorphic to that analyzed in
this section, and the two constructed quantizations equivalent.
In particular, the prediction that (for normalizable physical states)
the effective gravitational constant observed in the asymptotic regions of
large volume is positive [11] remains still valid in our quantization.

\section {KS Model: Quantization}
\setcounter{equation}{0}

Let us study now the other minisuperspace model for which we know a complete
set of quantum wormhole solutions, i.e., a KS spacetime minimally coupled to
an homogeneous massless scalar field $\Phi$ [8,17]. The KS metric can be
expressed in the form
\begin{equation} ds^2=\frac{G}{2\pi}\left(-q^4(t) e^{-2\alpha(t)}N^2(t)
dt^2+e^{2\alpha(t)}dr^2+q^2(t)e^{-2\alpha(t)}d\Omega_2^2\right),
\end{equation}
where $d\Omega_2^2$ is the metric of the unit two-sphere, $r$ is a periodic
coordinate, with period equal to $2\pi$, $N(t)$ is the rescaled lapse
function, and $e^{\alpha(t)}$ and $q(t)$ are the two scale factors of the
model. As in the case of the FRW spacetime, we will restrict the scale
factors to take only positive values, so that each Lorentzian metric is
considered only once. Thus, we will impose $q>0$ and $\alpha\in
I\!\!\!\,R$.

The model is subject only to a first-class constraint, provided by the
Hamiltonian of the system [17],
\begin{equation} H\equiv \frac{1}{2}\left(q^2\Pi_q^2+q^2
-\Pi_{\alpha}^2-\Pi_{\chi}^2\right)=0\;\;\;\;(q>0).\end{equation}
Here, $\Pi_q$, $\Pi_{\alpha}$, and $\Pi_{\chi}$ are the momenta canonically
conjugate to $q$, $\alpha$, and
\begin{equation} \chi(t)=\sqrt{4\pi G}\;\Phi(t),\end{equation}
respectively. For Lorentzian gravity and real fields we will have
$\chi,\Pi_q,\Pi_{\alpha},\Pi_{\chi}\in I\!\!\!\,R$.

The canonical transformation defined by
\begin{equation}
q=\frac{\Pi}{\cosh{X}},\;\;\;q\Pi_q=-\Pi\tanh{X},\end{equation}
\begin{equation} \chi=\phi\sin{\theta}+\frac{z}{\Pi_{\phi}}\cos{\theta},\;\;\;
\Pi_{\chi}=\Pi_{\phi}\sin{\theta},\end{equation}
\begin{equation} \alpha=\phi\cos{\theta}-\frac{z}{\Pi_{\phi}}\sin{\theta},\;\;
\;\Pi_{\alpha}=\Pi_{\phi}\cos{\theta},\end{equation}
allows us then to rewrite the constraint (3.2) as
\begin{equation} H\equiv \frac{1}{2}(\Pi^2-\Pi_{\phi}^2)=0.\end{equation}
In the above equations, $X$, $\phi$, and $z$ are the new configuration
variables and $\Pi$, $\Pi_{\phi}$, and $\theta$ their respective canonically
conjugate momenta. A generating function for the introduced transformation is
given by
\begin{equation} F(q,\alpha,\chi,X,\Pi_{\phi},\theta)=-q\sinh{X}+
\Pi_{\phi}(\alpha \cos{\theta}+\chi \sin{\theta}).\end{equation}
Recalling that $q>0$, it is not difficult to check from
eqs. (3.4-6) that, in the new set of variables, the domains
\begin{equation} X,\phi,z\in I\!\!\!\,R,\;\;\;\Pi,\Pi_{\phi}>0,\;\;\; \theta
\in S^1,\end{equation}
cover the whole phase space of the model, except for those points with
$\Pi_{\alpha}=\Pi_{\chi}=0$ ($\Pi_{\phi}=0$), that can never be reached in the
physical solutions owing to the Hamiltonian constraint (3.7).
The canonical transformation (3.4-6) turns out then to be
analytic and invertible in the region (3.9).

Notice that, from eq. (3.9), the phase space of the system can be regarded as
the cotangent bundle over $I\!\!\!\,R^+\times I\!\!\!\,R^+\times S^1$.
The inclusion of the points
with $\Pi_{\phi}=0$ would have not altered this result, because such points
are just part of the boundary of the considered phase space. On the other
hand, taking into account the Hamiltonian constraint (3.7), it is
straightforward to see that the reduced phase space of the
model is given by
the cotangent bundle over $I\!\!\!\,R^+\times S^1$ ($\Pi_{\phi}\in
I\!\!\!\,R^+$, $\theta\in S^1$).

To quantize the theory, we will select as representation space the vector
space of complex functions $\Psi(X,\Pi_{\phi},\theta)$ on $I\!\!\!\,R\times
I\!\!\!\,R^+\times S^1$, and define the following set of elementary operators
(with $\hbar=1$):
\begin{equation}\hat{\Pi}\Psi=-i\partial_X\Psi(X,\Pi_{\phi},\theta),
\;\;\;\hat{X}\Psi=X\,\Psi(X,\Pi_{\phi},\theta),\end{equation}
\begin{equation}\hat{\Pi}_{\phi}\Psi=\Pi_{\phi}\Psi(X,\Pi_{\phi},\theta),
\;\;\;\widehat{\Pi_{\phi}\phi}\Psi=i\left(\Pi_{\phi}\partial_{\Pi_{\phi}}
-\frac{1}{2}\right)\Psi(X,\Pi_{\phi},\theta),\end{equation}
\begin{equation} \widehat{\cos{\theta}}\Psi=\cos{\theta}\,\Psi(X,\Pi_{\phi},
\theta),\;\;\;\widehat{\sin{\theta}}\Psi=\sin{\theta}\,\Psi(X,\Pi_{\phi},
\theta),\end{equation}
\begin{equation}\hat{z}\Psi=i\partial_{\theta}\Psi(X,\Pi_{\phi},\theta).
\end{equation}
The action of $\widehat{\Pi_{\phi}\phi}$ in (3.11) has been chosen in a
convenient way to simplify the rest of our calculations.
The above set of operators forms a closed algebra [18] under commutation
relations:
\begin{equation} [\hat{X},\hat{\Pi}]=i,\;\;\;[\hat{\Pi}_{\phi},
\widehat{\Pi_{\phi}\phi}]=-i\hat{\Pi}_{\phi},\end{equation}
\begin{equation} [\hat{z},\widehat{\cos{\theta}}]=-i\,\widehat{\sin{\theta}},
\;\;\;[\hat{z},\widehat{\sin{\theta}}]=i\,\widehat{\cos{\theta}},
\end{equation}
all other commutators in the algebra being equal to zero.
In the quantum theory, the operators (3.10-13) represent the classical
variables $(\Pi,X,\Pi_{\phi},\Pi_{\phi}\phi,\cos{\theta},\sin{\theta},z)$,
which can be seen to provide a complete set of functions on
the phase space of the system. Recalling eq.
(3.9), one can easily show then that, for Lorentzian gravity,
the $\star$-involution induced
in the algebra of elementary operators (3.10-13) by the complex conjugation
relations between classical variables coincides with the identity operation.

Given the constraint (3.7) and that $\Pi,\Pi_{\phi}>0$, the physical states
of the model must satisfy the equation
\begin{equation}\hat{\Pi}\Psi(X,\Pi_{\phi},\theta)=\hat{\Pi}_{\phi}
\Psi(X,\Pi_{\phi},\theta),\end{equation}
whose solutions adopt the generic expression
\begin{equation}\Psi(X,\Pi_{\phi},\theta)=e^{iX\Pi_{\phi}}f(\Pi_{\phi},
\theta),\end{equation}
with $f(\Pi_{\phi},\theta)$ a complex function on $I\!\!\!\,R^+\times S^1$.

Since the reduced phase space of the system is the cotangent bundle over
$I\!\!\!\,R^+\times S^1$, a complete set of observables is provided by the
operators $\widehat{\cos{\theta}}$, $\widehat{\sin{\theta}}$, $\hat{z}$,
$\hat{\Pi}_{\phi}$ and [12]
\begin{equation} \hat{w}=\widehat{\Pi_{\phi}\phi}+\hat{\Pi}_{\phi}^2
\left(\hat{X}\hat{\Pi}^{-1}+\frac{i}{2}\hat{\Pi}^{-2}\right).\end{equation}
This last observable corresponds to the classical variable $\Pi_{\phi}\phi+
\Pi_{\phi}^2X\Pi^{-1}$. The term in brackets in (3.18) is simply the
symmetrized product of $\hat{X}$ and $\hat{\Pi}^{-1}$.
Using eqs. (3.10,11) and (3.16) it is possible to check that the operator
$\hat{w}$ acts in the following way on the physical states\footnote{In fact,
the definition of $\widehat{\Pi_{\phi}\phi}$ in (3.11) was chosen to
guarantee that eq. (3.19) holds.}
\begin{equation} \hat{w}\Psi(X,\Pi_{\phi},\theta)=e^{iX\Pi_{\phi}}\;i
\Pi_{\phi}\partial_{\Pi_{\phi}}f(\Pi_{\phi},\theta).\end{equation}

The inner product in the space of physical states can then be determined
by requiring the self-adjointness of the above set of observables, because
they all represent classical variables that are real in the Lorentzian
theory. In this way, one arrives at the inner product
\begin{equation} <\Upsilon,\Psi>=\int_{I\!\!\!\,R^+}\frac{d\Pi_{\phi}}
{\Pi_{\phi}}\int_{S^1}d\theta\, h^{\ast}(\Pi_{\phi},\theta)f(\Pi_{\phi},
\theta),\end{equation}
where $\Upsilon=e^{iX\Pi_{\phi}}h(\Pi_{\phi},\theta)$. Thus, the Hilbert
space of the studied Lorentzian model consists of all functions (3.17) with
$f(\Pi_{\phi},\theta)\in L^2(I\!\!\!\,R^+\times
S^1,d\Pi_{\phi}d\theta/\Pi_{\phi})$.

\section {KS Model: Quantum Wormholes}
\setcounter{equation}{0}

In order to discuss the role played in the quantum theory by the wormhole
states, we will translate now the results of the previous section
into the $(q,\alpha,\chi)$ representation that has been employed in the
literature to find the wormhole wave functions of the considered
KS model [8,17]. We expect that (at least for the physical states)
the change to such a representation be given by an integral
transform $\sigma$ of the type [12,19]:
\[\Psi(q,\alpha,\chi)=\sigma\left(\Psi(X,\Pi_{\phi},\theta)
\right)\]
\begin{equation}
=\int_{S^1}d\theta\int_{I\!\!\!\,R^+}d\Pi_{\phi}\int_{I\!\!\!\,R}
dX\,g(X,\Pi_{\phi},\theta)e^{iF(q,\alpha,\chi,X,\Pi_{\phi},\theta)}
e^{iX\Pi_{\phi}}f(\Pi_{\phi},\theta),\end{equation}
with $F$ the generating function (3.8), and $g$ a certain function on
$I\!\!\!\,R\times I\!\!\!\,R^+\times S^1$. Notice that on the right hand side
of (4.1) we have substituted the explicit form of the physical states (3.17).

To fix the function $g$ that appears in the integral transform $\sigma$ we
will demand the following conditions:
\begin{equation}\sigma\left(\hat{\Pi}\frac{1}{\cosh{X}}\Psi(X,\Pi_{\phi},
\theta)\right)\equiv
\hat{q}\Psi=q\,\Psi(q,\alpha,\chi),\end{equation}
\begin{equation}\sigma\left(-\hat{\Pi}\tanh{X}\Psi(X,\Pi_{\phi},\theta)\right)
\equiv\widehat{q\Pi_q}\Psi=
-iq\partial_q\Psi(q,\alpha,\chi),\end{equation}
\begin{equation}\sigma\left(\Pi_{\phi}\cos{\theta}\Psi(X,\Pi_{\phi},\theta)
\right)\equiv\hat{\Pi}_{\alpha}\Psi=
-i\partial_{\alpha}\Psi(q,\alpha,\chi),\end{equation}
\begin{equation} \sigma\left(\Pi_{\phi}\sin{\theta}\Psi(X,\Pi_{\phi},\theta)
\right)\equiv
\hat{\Pi}_{\chi}\Psi=-i\partial_{\chi}\Psi(q,\alpha,\chi),\end{equation}
\begin{equation} \sigma\left((\widehat{\Pi_{\phi}\phi}\cos^2{\theta}
-G(\theta,\hat{z}))\Psi(X,\Pi_{\phi},\theta)\right)\!\equiv
\widehat{\alpha\Pi_{\alpha}}\Psi=\!-i\left(\alpha\partial_
{\alpha}+\frac{1}{2}\right)\Psi(q,\alpha,\chi),\end{equation}
\begin{equation} \sigma\left((\widehat{\Pi_{\phi}\phi}\sin^2{\theta}+G
(\theta,\hat{z}))\Psi(X,\Pi_{\phi},\theta)\right)\!\equiv
\widehat{\chi\Pi_{\chi}}\Psi=\!-i\left(\chi\partial_
{\chi}+\frac{1}{2}\right)\Psi(q,\alpha,\chi),\end{equation}
where we have introduced the notation
\begin{equation} G(\theta,\hat{z})\Psi(X,\Pi_{\phi},\theta)=
\frac{1}{2}\left(\cos{\theta}\sin{\theta}\hat{z}+\hat{z}\cos{\theta}
\sin{\theta}\right)\Psi(X,\Pi_{\phi},\theta)\end{equation}
for the symmetrized product of the functions
$\cos{\theta}$ and $\sin{\theta}$ with the operator $\hat{z}$.

On the left hand side of eqs.(4.2-7) we have used the classical relations
(3.4-6) to define $(\hat{q},\widehat{q\Pi_q},
\hat{\Pi}_{\alpha},
\widehat{\alpha\Pi_{\alpha}},\hat{\Pi}_{\chi},\widehat{\chi\Pi_{\chi}})$ as
functions of the elementary operators (3.10-13)\linebreak
in the $(X,\Pi_{\phi},\theta)$
representation. In addition, the right hand sides of eqs. (4.2) and
(4.4,5) provide the standard action of the operators
$(\hat{q},\hat{\Pi}_{\alpha},\hat{\Pi}_{\chi})$ in the $(q,\alpha,\chi)$
representation. The action of $\widehat{\alpha\Pi_{\alpha}}$ in eq. (4.6), on
the other hand, can be regarded as that of the symmetrized product of
$\hat{\Pi}_{\alpha}$ and the multiplicative operator $\hat{\alpha}$
(with $\hat{\alpha}\Psi=\alpha\Psi(q,\alpha,\chi)$).
Similar considerations apply to the definition
of $\widehat{\chi\Pi_{\chi}}$ in eq. (4.7). Finally, the action of
$\widehat{q\Pi_q}$ and the factor ordering on the left hand side of eqs.
(4.2,3) have been chosen in such a way that, under the change of
representation, the quantum version of the Hamiltonian constraint (3.7)
translates into [12]
\begin{equation}\hat{H}\Psi(q,\alpha,\chi)=\frac{1}{2}\left(q^2-(q\partial_q)
^2+\partial_{\alpha}^2+\partial_{\chi}^2\right)\Psi(q,\alpha,\chi)=0,
\end{equation}
The integral transform $\sigma$ maps then physical states in
the $(X,\Pi_{\phi},\theta)$ representation into solutions to eq. (4.9),
which is precisely the Wheeler-DeWitt equation (obtained from constraint
(3.2)) for which we know explicitly a complete set of wormhole wave
functions.

Requirements (4.2-7) determine the function $g(X,\Pi_{\phi},\theta)$ in (4.1)
up to a constant:
\begin{equation} g=\frac{1}{\sqrt{\Pi_{\phi}}}.\end{equation}
With this choice for $g$, it is possible to prove that the change of
representation (4.1) is well-defined, at least, for all physical states
(3.17) in the Hilbert space of the model (i.e., with $f(\Pi_{\phi},\theta)
\in L^2(I\!\!\!\,R^+\times S^1,d\Pi_{\phi} d\theta/\Pi_{\phi})$). Substituting
then eqs. (3.8) and (4.10) in (4.1) and performing the integration over
the variable $X$ [12], we arrive at the following formula for the integral
transform $\sigma$:
\begin{equation} \Psi(q,\alpha,\chi)=\int_{S^1}d\theta\int_{I\!\!\!\,R^+}
\frac{d\Pi_{\phi}}{\sqrt{\Pi_{\phi}}}e^{i\Pi_{\phi}(\alpha\cos{\theta}+
\chi\sin{\theta})}2K_{i\Pi_{\phi}}(q)e^{\frac{\pi}{2}\Pi_{\phi}}f(\Pi_{\phi},
\theta),\end{equation}
where $K_{i\Pi_{\phi}}$ is a modified Bessel function of imaginary order [20].
This transformation can be inverted to regain the wave function
$f(\Pi_{\phi},\theta)$ that characterizes the physical states in the
$(X,\Pi_{\phi},\theta)$ representation [12]:
\begin{equation} f(\Pi_{\phi},\theta)=\int_{I\!\!\!\,R^+}dq \,R(\Pi_{\phi})
K_{i\Pi_{\phi}}(q)\int_{I\!\!\!\,R}d\alpha\int_{I\!\!\!\,R}d\chi
e^{-i\Pi_{\phi}(\alpha\cos{\theta}+\chi\sin{\theta})}\Psi(q,\alpha,\chi),
\end{equation}
\begin{equation} R(\Pi_{\phi})=\frac{2}{\pi^2}\Pi_{\phi}\sqrt{\Pi_{\phi}}
\cosh({\pi\Pi_{\phi}})e^{-\frac{\pi}{2}\Pi_{\phi}}.\end{equation}
Employing eqs. (3.20) and (4.12,13) one can also obtain the expression of
the inner product in the constructed $(q,\alpha,\chi)$ representation.

Given the form of the integral transform (4.11), it is clear that, in the
$(X,\Pi_{\phi},\theta)$ representation, all the physical states
can be interpreted as superpositions of the wave functions
\begin{equation} \Psi_{\tilde{\Pi}_{\phi},\tilde{\theta}}= 2
K_{i\tilde{\Pi}_{\phi}}(q)e^{\frac{\pi}{2}\tilde{\Pi}_{\phi}}e^{i\tilde{\Pi}_
{\phi}(\alpha\cos{\tilde{\theta}}+\chi\sin{\tilde{\theta}})},\end{equation}
with $\tilde{\Pi}_{\phi}\in I\!\!\!\,R^+$ and $\tilde{\theta}\in S^1$.
These wave functions were discovered (apart from an overall constant
factor) by Campbell and Garay [17], and provide a complete set of wormhole
states for the model. They are asymptotically damped for large scale factors
($q\rightarrow \infty$) and regular for non-degenerate three-geometries
($q>0$). However, they are singular when the three-geometry collapses to zero
($q\rightarrow 0$), so that, in a rigorous way, they can only be
considered as generalized quantum wormhole solutions.

In conclusion, what we have shown is that the Hilbert space of physical
states determined in the $(q,\alpha,\chi)$ representation by the Lorentzian
reality conditions is spanned by the basis of generalized wormholes (4.14).
In this sense, the Hilbert space of the Lorentzian theory coincides with the
Hilbert space of wormhole states in the adopted representation.

Using eqs. (4.12,13) it is not difficult to check that the generalized
wormholes (4.14) are described in the $(X,\Pi_{\phi},\theta)$ representation
by the wave functions
\begin{equation} f(\Pi_{\phi},\theta)=\sqrt{\Pi_{\phi}}\delta(\Pi_{\phi}
-\tilde{\Pi}_{\phi})\delta(\theta-\tilde{\theta})\equiv f_{\tilde{\Pi}_{\phi},
\tilde{\theta}}(\Pi_{\phi},\theta),\end{equation}
$\delta(\theta)$ being the delta function on $S^1$.
The wormhole solutions (4.14) are then mutually orthogonal with respect to
the physical inner product (3.20), although none of them is normalizable:
\begin{equation} <\Psi_{\tilde{\Pi}_{\phi}',\tilde{\theta}'},
\Psi_{\tilde{\Pi}_{\phi},\tilde{\theta}}>=\delta(\tilde{\Pi}_{\phi}'
-\tilde{\Pi}_{\phi})\delta(\tilde{\theta}'-\tilde{\theta}).\end{equation}
Employing eqs. (4.11) and (4.14), on the other hand, we can rewrite the wave
functions in the $(q,\alpha,\chi)$ representation as
\begin{equation} \Psi(q,\alpha,\chi)=\int_{S^1}d\tilde{\theta}
\int_{I\!\!\!\,R^+}d\tilde{\Pi}_{\phi}\bar{f}(\tilde{\Pi}_{\phi},
\tilde{\theta})\Psi_{\tilde{\Pi}_{\phi},\tilde{\theta}}(q,\alpha,\chi),
\end{equation}
where $\bar{f}(\tilde{\Pi}_{\phi},\tilde{\theta})=f(\tilde{\Pi}_{\phi},
\tilde{\theta})/\sqrt{\tilde{\Pi}_{\phi}}$ is the contribution of the quantum
wormhole $\Psi_{\tilde{\Pi}_{\phi},\tilde{\theta}}$ to the physical state
$\Psi$. The inner product (3.20) then takes the following form:
\begin{equation} <\Upsilon,\Psi>=\int_{S^1}d\theta\int_{I\!\!\!\,R^+}d
\Pi_{\phi} \bar{h}^{\ast}(\Pi_{\phi},\theta)\bar{f}(\Pi_{\phi},\theta)
\end{equation}
(with $\bar{h}=h/\sqrt{\tilde{\Pi}_{\phi}}$). As a consequence, the function
$|\bar{f}(\tilde{\Pi}_{\phi},\tilde{\theta})|^2$ can be interpreted, for
states of unit norm, as the probability to find a wormhole with parameters
$\tilde{\Pi}_{\phi}$ and $\tilde{\theta}$, the sum of all probabilities being
equal to one.

Finally, we notice that the parameters $(\tilde{\Pi}_{\phi},\tilde{\theta})$
that describe the different wormhole wave functions (4.15) (or, equivalently,
(4.14)) are determined by the eigenvalues reached in such solutions by the
complete set of compatible observables
$\{\hat{\Pi}_{\phi},\widehat{\cos{\theta}},\widehat{\sin{\theta}}\}$.

\section {KS Model: Other Wormhole Wave Functions}
\setcounter{equation}{0}

We now wish to consider other known sets of wormhole wave functions
for the KS model under analysis.

One of these sets was first found by Campbell and Garay
[17], and its elements can be obtained as Euclidean path integrals over
manifolds with asymptotic large three-volume and no gravitational excitations
[8]:
\begin{equation} \Psi_{\tilde{\theta},\alpha_0,\chi_0}(q,\alpha,\chi)=
e^{-q\cosh{[(\alpha-\alpha_0)\cos{\tilde{\theta}}+(\chi-\chi_0)
\sin{\tilde{\theta}}\,]}}\;.\end{equation}
Here, $\alpha_0$ and $\chi_0$ run over the real axis, and coincide
with the asymptotic values of the fields $\alpha$ and $\chi$, respectively,
in the region of large three-volume [8]. On the other hand, given that the
parameters $\tilde{\theta}$ and $\tilde{\theta}+\pi$ lead to the same wave
function in (5.1), we will restrict $\tilde{\theta}\in [0,\pi)$ from now on.

The physical states (5.1) are obviously regular for all finite
fields $\alpha$ and $\chi$ and non-negative scale factors $q$
(including $q=0$). They are also exponentially damped when the scale factor
tends to infinity. Thus, they can be considered truly wormhole solutions of
the model.

A careful calculation using formulas (4.12,13) shows that, in the
$(X,\Pi_{\phi},\theta)$ representation constructed in Sec. III, these
wormhole states are described by the wave functions:
\[ f(\Pi_{\phi},\theta)=\frac{\sqrt{\Pi_{\phi}}}{2\pi}e^{-\frac{
\pi}{2}\Pi_{\phi}}e^{-i\Pi_{\phi}(\alpha_0\cos{\theta}+\chi_0\sin{\theta})}
\,[\delta(\theta-\tilde{\theta})+\delta(\theta-\tilde{\theta}-\pi)]\]
\begin{equation} \equiv f_{\tilde{\theta},\alpha_0,\chi_0}(\Pi_{\phi},\theta).
\end{equation}
This result can be easily checked by substituting eq. (5.2) in (4.11)
and employing that [21]
\begin{equation} \int_{I\!\!\!\,R}dp\,e^{-ip\omega}K_{ip}(q)=\pi
e^{-q\cosh{\omega}}\;\;\;\; {\rm for} \;q>0.\end{equation}
{}From eq. (5.2) and the form of the physical inner product (3.20) it
is then straightforward to conclude that the wormhole wave functions
(5.1) do not correspond to normalizable physical states.

Another set of solutions to the constraint (4.9) is provided by [17]
\begin{equation} \Psi_{n,\tilde{\theta}}(q,\alpha,\chi)=\varphi_n(u)\varphi_n
(v)=\frac{1}{2^n n!\sqrt{\pi}}H_n(u)H_n(v)e^{-q\cosh{(\alpha
\cos{\tilde{\theta}}+\chi\sin{\tilde{\theta}})}},\end{equation}
\begin{equation} u=\sqrt{2q}\cosh{\!\left(\frac{\alpha\cos{\tilde{\theta}}+
\chi\sin{\tilde{\theta}}}{2}\right)},\;\;\;v=\sqrt{2q}\sinh{\!\left(
\frac{\alpha\cos{\tilde{\theta}}+\chi\sin{\tilde{\theta}}}{2}\right)},
\end{equation}
where $\varphi_n$ are the harmonic-oscillator wave functions (2.5), $H_n$ the
Hermite polynomials [21], and $n$ is a non-negative integer ($n=0,1,...$).
Given the parity properties of the Hermite polynomials, the quantum states
$\Psi_{n,\tilde{\theta}}$ and $\Psi_{n,\tilde{\theta}+\pi}$ differ just by a
factor of $(-1)^n$. In order to consider only linearly independent
solutions, it will thus suffice to analyze the sector $\tilde{\theta}\in
[0,\pi)$.

The above wave functions represent proper quantum wormholes, because they
are regular everywhere in the configuration space
$\{q>0,\;\,\alpha,\chi\!\in \!I\!\!\!\,R\}$,
decrease exponentially for large values of the scale factor
$q$, and have a well-defined limit when the three-geometry degenerates (i.e.,
when $q\rightarrow 0$). Moreover, they actually form a basis for the Hilbert
space of the Lorentzian model, for all the elements of the basis (4.14) can be
expressed as superpositions of the wormhole solutions (5.4,5) [11,12]:
\begin{equation} \Psi_{\tilde{\Pi}_{\phi},\tilde{\theta}}(q,\alpha,\chi)=
\sum_{n=0}^{\infty} F_n(\tilde{\Pi}_{\phi}) \Psi_{n,\tilde{\theta}}(q,\alpha,
\chi),\end{equation}
\begin{equation} F_n(\tilde{\Pi}_{\phi})=2\sqrt{\pi}e^{\frac{\pi}{2}
\tilde{\Pi}_{\phi}}\int_{I\!\!\!\,R}d\eta e^{i2\tilde{\Pi}_{\phi}\eta}
\,\frac{\sinh^n{\eta}}{\cosh^{n+1}{\eta}}.\end{equation}
We conclude in this way that the Hilbert space of the constructed
quantum theory admits indeed a basis of wormhole states, in agreement with our
comments in the Introduction.

The wave functions (3.17) that correspond to the wormholes
$\Psi_{n,\tilde{\theta}}$ in the $(X,\Pi_{\phi},\theta)$ representation
can be obtained using eqs. (4.12,13):
\begin{equation}
f(\Pi_{\phi},\theta)=\!\frac{\sqrt{\Pi_{\phi}}}{2\pi^2}e^{-\pi
\Pi_{\phi}}F_n^{\ast}(\Pi_{\phi})[\delta(\theta-\tilde{\theta})+(-1)^n\delta
(\theta-\tilde{\theta}-\pi)]\equiv\! f_{n,\tilde{\theta}}
(\Pi_{\phi},\theta),\end{equation}
with $\tilde{\theta}\in [0,\pi)$. It is then easy to check that these wormhole
solutions are not normalizable with respect to the physical inner
product (3.20).

{}From the wormhole wave functions (5.4,5), nevertheless, one can arrive at
a new basis of states whose elements have all finite norm, as we will show
in the rest of this section. Let us first substitute relation (5.6) (and
(4.14)) in eq. (4.11). Assuming that we
can interchange the order of summation and integration,
we get the following expression for all physical states
in the Hilbert space of the model:
\begin{equation} \Psi(q,\alpha,\chi)=\sum_{n=0}^{\infty}\int_0^{\pi}
d\theta \,2 h_n(\theta)\Psi_{n,\theta}(q,\alpha,\chi),\end{equation}
where
\begin{equation}h_n(\theta)=\int_{I\!\!\!\,R^+}\frac{d\Pi_{\phi}}{2\sqrt{
\Pi_{\phi}}}F_n(\Pi_{\phi})\left[f(\Pi_{\phi},\theta)+(-1)^n
f(\Pi_{\phi},\theta+\pi)\right].\end{equation}
{}From eqs. (5.4,5) and (5.10), it is not difficult to prove that the functions
\begin{equation} \tilde{\Psi}_{n,\theta}(q,\alpha,\chi)=e^{in\theta}
\Psi_{n,\theta}(q,\alpha,\chi),\;\;\;\tilde{h}_n(\theta)=e^{-in\theta}
h_n(\theta)\end{equation}
are periodic\footnote{We recall that $f(\Pi_{\phi},\theta)$ is defined
on $I\!\!\!\,R^+\times S^1$.} in $\theta$, with period equal to
$\pi$. Therefore, they both admit a Fourier expansion in the unit circle
described by the angle $\beta=2\theta$. As a consequence, eq. (5.9) can be
rewritten in the form
\begin{equation} \Psi(q,\alpha,\chi)=\sum_{n=0}^{\infty}\sum_{m=-\infty}
^{\infty}\tilde{h}_{nm}\tilde{\Psi}_{nm}(q,\alpha,\chi),\end{equation}
with
\begin{equation}\tilde{h}_{nm}=\int_{S^1}\frac{d\beta}{\sqrt{2\pi}}
e^{-im\beta}\tilde{h}_n\left(\frac{\beta}{2}\right),\end{equation}
\begin{equation}\tilde{\Psi}_{nm}(q,\alpha,\chi)=\int_{S^1}\frac{d\beta}
{\sqrt{2\pi}}e^{im\beta}\tilde{\Psi}_{n,\beta/2}(q,\alpha,\chi).\end{equation}
Eq. (5.12) implies that all the normalizable physical states are
superpositions of the wave functions $\{\tilde{\Psi}_{nm},\;n=0,1,...,\;
m=0,\pm 1,...\}$, which thus provide a discrete basis for the Hilbert
space of the Lorentzian model.

To find the expression of the states $\tilde{\Psi}_{nm}$ in the
$(X,\Pi_{\phi},\theta)$ representation one can use eqs. (4.12,13), (5.4,5),
(5.11) and (5,14). In fact, that amounts to replace the function
$\tilde{\Psi}\,\!_{n,\beta\,\!\!/2}$$(q,\alpha,\chi)$
on the right hand side of (5.14)
by $e^{in\beta/2}f_{n,\beta/2}(\Pi_{\phi},\theta)$ (with $f_{n,\beta/2}$
given by eq. (5.8)) and perform then the integration over $\beta$. In doing
so, one arrives at the result:
\begin{equation}f(\Pi_{\phi},\theta)=e^{i(2m+n)\theta}\frac{\sqrt{\Pi_{\phi}}}
{\pi^2\sqrt{2\pi}}\cosh{(\pi\Pi_{\phi})}e^{-\pi\Pi_{\phi}}F_n^{\ast}
(\Pi_{\phi})\equiv f_{nm}(\Pi_{\phi},\theta),\end{equation}
and, from eq. (3.17),
\begin{equation} \tilde{\Psi}_{nm}(X,\Pi_{\phi},\theta)=e^{iX\Pi_{\phi}}
f_{nm}(\Pi_{\phi},\theta).\end{equation}
Substituting now eq. (5.15) in the formula of the inner product (3.20),
we get that
\begin{equation}<\tilde{\Psi}_{pq},\tilde{\Psi}_{nm}>=\delta_{p+2q,n+2m}
\int_{I\!\!\!\,R^+}\frac{d\Pi_{\phi}}{\pi^4}\cosh^2(\pi\Pi_{\phi})
e^{-2\pi\Pi_{\phi}}F_p(\Pi_{\phi})F_n^{\ast}(\Pi_{\phi}),\end{equation}
with $\delta_{nm}$ the Kronecker delta. The integral over $\Pi_{\phi}$
in the above equation
converges for all possible values of $p$ and $n$, because the integrand
is continuous $\forall\Pi_{\phi}\in (0,\infty)$, has a finite limit when
$\Pi_{\phi}$ tends to zero, and decrease faster than $\Pi_{\phi}^{-1}$ at
infinity. This last statement follows from the form of the function $F_n
(\Pi_{\phi})$ (5.7) and the fact that [12]
\begin{equation} \left|\lim_{\Pi_{\phi}\rightarrow\infty}e^{\frac{\pi}{2}
\Pi_{\phi}}\Pi_{\phi}^2\int_{I\!\!\!\,R}d\eta e^{\pm i2\Pi_{\phi}\eta}\,
\frac{\sinh^n{\eta}}{\cosh^{n+1}{\eta}}\right|<\infty.\end{equation}
Therefore, the wave functions $\tilde{\Psi}_{nm}$ turn out to be
normalizable with respect to the physical inner product (3.20). Since the
basis $\{\tilde{\Psi}_{nm}\}$ is discrete, we conclude that it can be
orthonormalized following the standard Gram-Schmidt technique. Finally,
we notice that the Hilbert space of the considered Lorentzian model is then
separable, for it admits a discrete basis of orthonormal states.

\section {Conclusions}

We have analyzed the Hilbert structure of the space of wave functions spanned
by the quantum wormhole solutions in a FRW model provided with
a conformally coupled scalar field and in a KS spacetime minimally coupled
to a massless scalar field. We have proved that the inner product in such
a space can be uniquely fixed by imposing a set of Lorentzian reality
conditions. The Hilbert spaces determined in this way have been shown to
coincide, for the adopted representations, with those of the respective
Lorentzian models under study. This implies in particular that all the
normalizable physical states can be interpreted in the Lorentzian theory
as superpositions of wormholes.

The results of this work, together with those presented in Ref. [12]
(where the case of a FRW spacetime with a minimally coupled scalar field was
considered), prove that, in all the minisuperspace models in which a complete
set of wormhole wave functions is known, one can carry out to completion the
quantization of the Lorentzian theory following Ashtekar's program, and that,
with an adequate choice of representation, the obtained Hilbert space turns
out to admit a basis of quantum wormhole states. In all these models,
we have seen in addition that it is possible to find a discrete orthonormal
basis of wormholes, so that the Hilbert spaces of the corresponding Lorentzian
systems are separable.

For Lorentzian General Relativity, we expect that the Hilbert space of
wormholes can be identified as the physical Hilbert space if: 1) one
can find a representation for full gravity in which there exist wormhole wave
functions, and 2) the vector space spanned by those wave functions is
stable under the action of the gravitational observables. If this is the case,
one can choose the vector space of wormholes as the (restricted)
representation space for the quantum theory of gravity and complete
the quantization by
demanding an appropriate set of reality conditions. By construction, the
Hilbert space of physical states will then be spanned by a basis of wormholes.
We notice that such a quantization will be meaningful only if the space of
wormholes is neither empty nor trivial, therefore the first condition.
The second condition, on the other hand, guarantees that the reality
conditions on the observables can be imposed as adjointness relations
in the Hilbert space of wormhole solutions.

{\bf Acknowledgements}

\vspace*{.4cm}
The author is grateful to P. Gonz\'alez D\'{\i}az and D. Marolf for
helpful conversations. He also wants to thank the Center for Gravitational
Physics and Geometry at the Pennsylvania State University for warm
hospitality.


\newpage

\begin{thebibliography}{22}

\bibitem{1} S. W. Hawking, Phys. Rev. D {\bf 37}, 904 (1988).

\bibitem{2} S. Giddings and A. Strominger, Nucl. Phys. {\bf B306}, 890 (1988).

\bibitem{3} J. J. Halliwell and R. Laflamme, Class. Quantum Grav. {\bf 6},
1839 (1989).

\bibitem{4} A. Hosoya and W. Ogura, Phys. Lett. B {\bf 225}, 117 (1989).

\bibitem{5} S. Bochner, Bull. Am. Math. Soc. {\bf 52}, 776 (1946).

\bibitem{6} J. Cheeger and D. Gromoll, Ann. Math. {\bf 96}, 413 (1972).

\bibitem{7} S. W. Hawking and D. N. Page, Phys. Rev. D {\bf 42}, 2655 (1990).

\bibitem{8} L. J. Garay, Phys. Rev. D {\bf 44}, 1059 (1991).

\bibitem{9} S. Coleman, Nucl. Phys. {\bf B310}, 643 (1988).

\bibitem{10} S. W. Hawking, Nucl. Phys. {\bf B335}, 155 (1990).

\bibitem{11} L. J. Garay, Phys. Rev. D {\bf 48}, 1710 (1993).

\bibitem{12} G. A. Mena Marug\'an, ``Bases of Wormholes in Quantum
Cosmology'', Penn State University Report No. CGPG-94/4-4, gr-qc/9404042
(unpublished).

\bibitem{13} A. Ashtekar, {\it Lectures on Non-Perturbative Canonical
Gravity}, edited by L. Z. Fang and R. Ruffini (World Scientific, Singapore,
1991).

\bibitem{14} A. Ashtekar, in {\it Gravitation and Quantizations},
proceedings of the Les Houches Summer School, Les Houches, France, 1992,
edited by B. Julia and J. Zinn-Justin, Les Houches Summer School Proceedings
Vol. LVII (North Holland, Amsterdam, 1993).

\bibitem{15} G. A. Mena Marug\'an, ``Reality Conditions for Lorentzian and
Euclidean Gravity in the Ashtekar Formulation'', Penn State University
Report No. CGPG-93/11-2, Int. J. Mod. Phys. D (to be published).

\bibitem{16} A. Rendall, Class. Quantum Grav. {\bf 10}, 2261 (1993).

\bibitem{17} L. M. Campbell and L. J. Garay, Phys. Lett. B {\bf 254}, 49
(1991).

\bibitem{18} Together with the identity operator.

\bibitem{19} G. A. Mena Marug\'an, Class. Quantum Grav. {\bf 11}, 589 (1994).

\bibitem{20} M. Abramowitz and I. A. Stegun (eds.), {\it Handbook of
Mathematical Functions}, Natl. Bur. Stand. Appl. Math. Ser. No. 55, revised
9th. ed. (U.S. Govt. Print. Off., Washington, D.C., 1970).

\bibitem{21} I. S. Gradshteyn and I. M. Ryzhik, {\it Table of Integrals,
Series and Products}, revised 4th. ed. (Academic Press, San Diego, 1980).

\end{thebibliography}
\end{document}